\documentclass[11pt]{article}
\usepackage[margin=1in]{geometry}
\usepackage{amsmath,amssymb,booktabs,array,float}
\usepackage[T1]{fontenc}\usepackage{lmodern}\usepackage{hyperref}
\hypersetup{colorlinks=true,linkcolor=blue,citecolor=blue,urlcolor=blue}
\title{Flowz: When Does Learning Forecast Errors Add Value?\\Conditional Flow Matching for Financial Price Forecasts}
\author{Giancarlo Vercellino\thanks{Contact: \href{mailto:giancarlo.vercellino@gmail.com}{giancarlo.vercellino@gmail.com}.}}
\date{October 21, 2026}
\begin{document}\maketitle
\begin{abstract}
Learning a forecasting model's errors offers a modular way to improve its
predictive distribution while retaining its underlying dynamics. We study
this approach with Flowz, a conditional flow model of direct multi-step
price-level residuals fitted using random features and ridge regression.
Each EWMA, ARIMA, ARFIMA, and structural baseline is compared with a Gaussian
residual location--scale control and a conditional Flowz extension. The
experiment uses 24 US stocks selected across four liquidity--volatility
strata, 104 common forecast origins, horizons of 5, 10, 20, 40, 80, and 160
trading observations, and three fit seeds. Marginal CRPS is the primary score.
At farther horizons, Flowz reduces mean CRPS by 30.7\% against EWMA and 24.0\% against its Gaussian residual control.
The EWMA improvement holds at each fit seed and exceeds the Gaussian control for all 24 stocks in the far band. The ARIMA, ARFIMA, and structural extensions worsen both comparators. Conditional residual learning is therefore useful for the drifting EWMA specification in this experiment, with gains concentrated at longer horizons.
\end{abstract}

\section{Introduction}
Financial forecasts must describe both future price levels and the uncertainty
surrounding them. A forecast can have a useful central estimate but poorly
represent the range of subsequent outcomes. Conversely, wide intervals can
cover observations while providing little precision. Proper scoring rules
evaluate the predictive distribution as a whole \cite{gneiting}.

The question of how to improve these distributions is timely. Financial
stability analysis highlights the connection between uncertainty, market
functioning, and abrupt repricing \cite{imf2025}. At the same time, models such
as Chronos illustrate rapid advances in probabilistic time-series forecasting
\cite{chronos}. A complementary statistical route is to retain an existing
forecasting model and learn the structure of its errors. This route is
attractive when the baseline already captures useful dynamics and can be
updated economically as new observations arrive.

Flowz implements this approach by learning joint vectors of future price-level
errors conditional on a local market state. A baseline supplies the forecast
mean and marginal scales. Historical errors are centered and standardized,
and a conditional flow maps Gaussian inputs into residual scenarios. Adding
those scenarios to the baseline produces a predictive distribution for future
levels. The statistical value of this construction depends on the information
remaining in the baseline's errors.

We ask three research questions:
\begin{enumerate}
\item \textbf{Incremental value:} Does conditional residual learning improve
the predictive distribution of the same underlying baseline?
\item \textbf{Value beyond location and scale:} Does Flowz improve a Gaussian
control with the same training-only residual centering and scaling?
\item \textbf{Conditions of usefulness:} How do these effects vary across
forecast horizons, baseline engines, liquidity, and volatility?
\end{enumerate}
The working hypothesis is that conditional residual learning improves CRPS,
particularly at longer horizons where baseline errors accumulate. We examine
this hypothesis through matched comparisons across four baseline engines and
a recent stock panel. The experiment is exploratory: all engines and all three
fit seeds enter the reported results.

\section{Related methods}
ARIMA describes linear short-memory dynamics after integer differencing
\cite{box}; ARFIMA extends this representation through fractional differencing
\cite{hosking}. EWMA and structural state-space models provide alternative
local descriptions. They form four starting points for evaluating the
incremental effect of conditional residual learning.

Flow matching estimates velocity fields by regression along probability paths
\cite{lipman}. Flowz combines this construction with fixed random hidden
features and ridge regression, following the extreme learning machine approach
\cite{huang}. FlowTime uses autoregressive flow matching for forecasting
\cite{flowtime}; FreqFlow also uses residual learning, with a frequency-domain
formulation \cite{freqflow}. Flowz instead models a complete univariate vector
of future level errors directly, conditioned on local innovations, drift,
volatility, and recent increments.

The matched design connects this generative construction to a familiar
statistical question: how much predictive information remains after a baseline
has been fitted? A Gaussian location--scale control separates the effect of
unconditional residual centering and scaling from the additional conditional
distribution learned by the flow.

\section{Conditional residual forecasting with Flowz}
\subsection{Target and local state}
Let $P_t>0$ be the split-adjusted closing price and $y_t=\log P_t$.
At origin $i$, the target is the conditional distribution of
$\mathbf y_i^+=(y_{i+1},\ldots,y_{i+H})^\top$. We use $H=160$ and score six
endpoints of each generated vector. Horizons count trading observations.

Write $\Delta_t=y_t-y_{t-1}$. The local state is updated through
\begin{align}
e_t&=\Delta_t-\mu_{t-1},\\
\mu_t&=\lambda_\mu\mu_{t-1}+(1-\lambda_\mu)\Delta_t,\\
\sigma_t^2&=\lambda_\sigma\sigma_{t-1}^2+(1-\lambda_\sigma)e_t^2.
\end{align}
The implementation initializes $\mu_1=0$, estimates an initial scale from the
first \texttt{min\_train} observations, and uses positive variance floors.
Eligible historical origins begin at the end of this initial block.
For EWMA, the baseline mean and marginal standard deviation are
\begin{equation}
b_{i,h}=y_i+h\mu_i,\qquad s_{i,h}=\sigma_i\sqrt h.
\end{equation}
The context contains eight recent standardized innovations $e_t/\sigma_t$,
the ratio $\mu_i/\sigma_i$, $\log\sigma_i$, and mean increments over the latest
three and ten observations divided by $\sigma_i$. These 12 features are
centered and scaled using training origins.

\subsection{Baseline engines}
The four baseline options are \texttt{ewma}, \texttt{arima}, \texttt{arfima},
and \texttt{structural}. The parametric engines select a specification on the
first 200 observations. At subsequent origins they refit coefficients using
the history available through that origin, while retaining the selected order
or structural type. EWMA updates its local state recursively.

For ARIMA, \texttt{forecast::auto.arima} selects a nonseasonal model on log
levels using a stepwise search, exact likelihood evaluation, $p,q\leq3$, and
$d\leq1$ \cite{hyndman}. Subsequent fits use \texttt{forecast::Arima}, retaining
the selected order and inclusion of a mean or drift.

For ARFIMA, \texttt{forecast::arfima} selects the initial orders on log levels
with fractional range $[0,0.5]$ and maximum likelihood estimation. At later
origins, \texttt{fracdiff::fracdiff} refits those orders and
\texttt{forecast::arfima} supplies the forecast \cite{haslett,fracdiff}.
The fractional parameter is re-estimated at each origin.

For the structural engine, \texttt{stats::StructTS} selects between local
level and local linear trend by $-2\log L+2k$ on the initial window. The local
level model is $y_t=\ell_t+\epsilon_t$, $\ell_t=\ell_{t-1}+\xi_t$; local
linear trend adds a slope state to the level evolution. Observation and state
disturbance variances are estimated from the available history.

ARIMA and ARFIMA marginal scales are obtained from symmetric 90\% forecast
limits by dividing their width by $2\Phi^{-1}(0.95)$. Structural fits return
forecast standard errors directly. If a rolling fit or forecast raises an
error, the implementation uses the causal EWMA mean and scale at that origin.
All three compared outputs share this fallback policy.

For any engine, the baseline sample is
\begin{equation}
B_{i,h}^{(m)}=b_{i,h}+s_{i,h}\eta_{m,h},\qquad
\eta_{m,h}\stackrel{\mathrm{iid}}\sim N(0,1),\quad m=1,\ldots,M.
\end{equation}
Here $M=300$ for external evaluation. These draws use independent Gaussian
marginals across horizons. The baseline comparison therefore concerns marginal
forecast distributions; Flowz generates a joint residual vector.

\subsection{Direct residual vectors}
Let $\mathcal T$ be the training-origin set and define
$E_{i,h}=y_{i+h}-b_{i,h}$. The training mean error is
$a_h=|\mathcal T|^{-1}\sum_{i\in\mathcal T}E_{i,h}$. Define
\begin{equation}
k_h=\operatorname{SD}_{i\in\mathcal T}
\left(\frac{E_{i,h}-a_h}{s_{i,h}}\right),\qquad
R_{i,h}=\frac{E_{i,h}-a_h}{s_{i,h}k_h}.
\end{equation}
Numerical safeguards replace degenerate scale corrections. The vector
$\mathbf R_i\in\mathbb R^H$ contains standardized future level errors.
Centering $a_h$, scaling $k_h$, and context standardization are frozen after
training. Direct vector prediction represents dependence across horizons
without feeding generated observations back into the baseline model.

\subsection{Conditional flow matching}
For each training endpoint and augmentation draw, independently sample
$\mathbf z_0\sim N(\mathbf0,I_H)$ and $u\sim U(0,1)$. Set
\begin{equation}
\mathbf z_u=(1-u)\mathbf z_0+u\mathbf R_i,\qquad
\mathbf v_i=\mathbf R_i-\mathbf z_0.
\end{equation}
With standardized context $\widetilde{\mathbf c}_i$, the regression input is
$\mathbf x=(\mathbf z_u^\top,u,\widetilde{\mathbf c}_i^\top)^\top$.
Its dimension is $D=H+1+12=173$. Hidden features are
\begin{equation}
\phi_j(\mathbf x)=\tanh(\mathbf w_j^\top\mathbf x+b_j),\qquad
\mathbf w_j\sim N(\mathbf0,I_D/D),\quad b_j\sim U(-1,1).
\end{equation}
Random hidden weights and biases remain fixed. Let $A$ contain an intercept
and the hidden features, and let $V$ contain velocity targets. Ridge regression
estimates
\begin{equation}
\widehat B=(A^\top A+\Lambda)^{-1}A^\top V,\qquad
\Lambda=\operatorname{diag}(0,\lambda,\ldots,\lambda).
\end{equation}
The implementation solves the linear system with an unpenalized intercept.
Augmentation draws approximate the conditional squared-velocity regression
objective.

At a new origin, draw Gaussian inputs and solve
\begin{equation}
\frac{d\mathbf z_u}{du}
=\widehat{\mathbf v}(\mathbf z_u,u,\widetilde{\mathbf c}_i),
\qquad 0\leq u\leq1,
\end{equation}
using fourth-order Runge--Kutta integration. The forecast scenarios are
\begin{equation}
Y_{i,h}^{(m)}=b_{i,h}+a_h+s_{i,h}k_h Z_{1,h}^{(m)}.
\end{equation}
Transport time $u$ indexes the flow from Gaussian inputs to residuals;
financial time indexes observations and forecast horizons. Exponentiating
the generated log levels returns price scenarios.

\section{Matched experiment}
\subsection{Three outputs and their contrasts}
For each stock and baseline engine, we compare the baseline, the Gaussian
residual control, and Flowz, the conditional residual model. The control is
\begin{equation}
G_{i,h}^{(m)}=b_{i,h}+a_h+s_{i,h}k_h\eta_{m,h}
=b_{i,h}+a_h+k_h(B_{i,h}^{(m)}-b_{i,h}).
\end{equation}
It shares baseline Gaussian draws and uses the same training-only $a_h$ and
$k_h$ as Flowz. These parameters are unconditional summaries at each horizon.
The Flowz model additionally conditions the residual distribution on the local
state. The comparisons are organized as
\[
\text{baseline}\ \longrightarrow\ (a_h,k_h)\quad
\begin{cases}
\text{Gaussian residual control},\\
\text{conditional flow}\ \longrightarrow\ \text{Flowz scenarios}.
\end{cases}
\]
Flowz/base measures the full conditional residual pipeline; G/base measures
centering and scaling alone; Flowz/G measures the additional conditional model.
All outputs use the same history, baseline specification, forecast origins,
and number of simulations.

\subsection{Stock selection and price construction}
The panel contains 24 US stocks, six in each high/low liquidity by high/low
volatility cell. An Alpaca snapshot retrieved on October 9, 2026 contains
8,536 active NYSE/NASDAQ assets. A common-stock name filter leaves 2,511
names; the first 240 in a deterministic SHA-256 ordering are screened using
2016--2021 measurements. Of these, 81 meet the eligibility criteria: at least
98\% training-session coverage, at least 250 of the last 252 sessions,
median raw price of at least USD 5, and median daily dollar volume of at least
USD 100,000. Unsupported stock-dividend cases are excluded.

Liquidity is median raw closing price times reported volume over the final
252 sessions through December 31, 2021. Volatility is the sample standard
deviation of split-adjusted daily log returns, annualized by $\sqrt{252}$.
We retain the top and bottom liquidity quartiles (20 names each), then the
top and bottom volatility thirds within each quartile (six names per cell).
High/low labels are relative to the screened set. Selection uses these
pre-evaluation characteristics, independently of forecast losses.

\begin{table}[H]\centering\small
\caption{Selected stocks by pre-evaluation liquidity and volatility.}
\begin{tabular}{p{.40\textwidth}p{.51\textwidth}}\toprule Stratum & Symbols\\\midrule
high liquidity high volatility & AMAT, CAR, CLNE, CZR, FATE, RARE \\
high liquidity low volatility & CBRL, EXC, HAS, JBHT, SSNC, TROW \\
low liquidity high volatility & CHCI, GROW, JAKK, NGNE, PNRG, VRDN \\
low liquidity low volatility & CHMG, CZWI, GABC, GBLI, HBCP, YORW \\
\bottomrule\end{tabular}\label{tab:panel}\end{table}

\begin{table}[H]\centering\small
\caption{Medians of stock-level selection metrics across the six stocks in each cell. Liquidity is median daily dollar volume; volatility is annualized. Labels are relative to the 81 screened eligible names.}
\begin{tabular}{lrr}\toprule Stratum & USD million & Volatility (\%)\\\midrule
high liquidity high volatility & 106.799 & 61.5 \\
high liquidity low volatility & 94.784 & 24.7 \\
low liquidity high volatility & 0.731 & 71.5 \\
low liquidity low volatility & 0.505 & 27.3 \\
\bottomrule\end{tabular}\label{tab:strata}\end{table}

Daily closes span January 4, 2016--October 8, 2026: 2,707 exchange sessions,
including 1,511 through December 31, 2021. Forward split adjustment multiplies
each raw close by the cumulative new-to-old share ratio effective by that
date; predecessor symbols are reconciled for corporate actions. The target
is the split-adjusted price, excluding cash dividends and spin-off distributions.
All selected names have complete aligned coverage. Zero-volume and unchanged
closes are retained; Appendix~\ref{app:evidence} reports their counts.

\subsection{Chronology and forecast origins}
Each stock/engine fit uses an initial history ending December 31, 2021.
There are 300 candidate historical origins on a rounded equally spaced grid
from observation 200 to $1511-160=1351$. A chronological allocation of
65\%, 25\%, and 10\% defines the training block and two reserved historical
blocks. An embargo removes origins whose 160-step target extends beyond the
first origin of the next block, retaining 154, 34, and 30 origins respectively
and purging 82 candidates. Only the first block supplies residual-model
training; the other two blocks are excluded from this comparison's fitting
and reported scores. Retaining this allocation fixes the training sample.

\begin{table}[H]\centering\footnotesize
\caption{Chronological origin blocks and the final 160-observation target in each block.}
\begin{tabular}{lrll}\toprule Block & Origins & Origin dates & Final target\\\midrule
Residual training & 154 & 2016-10-17--2019-02-21 & 2019-10-09 \\
Reserved block 1 & 34 & 2019-10-11--2020-04-15 & 2020-12-01 \\
Reserved block 2 & 30 & 2020-12-02--2021-05-14 & 2021-12-31 \\
External evaluation & 104 & 2021-12-31--2026-02-10 & 2026-09-30 \\
\bottomrule\end{tabular}\label{tab:chronology}\end{table}

Training origins run from observations 200 to 789, with final targets at 949.
The reserved blocks run from origins 951--1078 and 1239--1351. The fitted
conditional residual model remains frozen throughout external evaluation.
The experiment thus studies a historical residual model augmenting a baseline
whose state or coefficients update as new data arrive.

External evaluation uses 104 common origins from December 31, 2021 to
February 10, 2026, spaced ten exchange sessions apart. The last 160-step target
ends September 30, 2026. Every output is scored at
$h\in\{5,10,20,40,80,160\}$ on these same origins. The near band comprises
5, 10, and 20 observations; the far band comprises 40, 80, and 160.

Settings are $\lambda_\mu=0.90$, $\lambda_\sigma=0.94$, eight innovation lags,
60 hidden units, five augmentations, ridge penalty $10^{-3}$, 20 RK4 steps,
and 300 external paths. Each stock/engine is fitted at seeds 123, 456, and
789. Forecast seeds are $10000+i$, held fixed across fits at origin $i$.
The three fits vary random features and augmentation draws while sharing
forecast random numbers. Residual centers and scale multipliers agree across
fit seeds. All fits use the same computational configuration.

\subsection{Scoring and paired uncertainty}
We evaluate origin-centered log levels $X_{i,h}=100(y_{i+h}-y_i)$ and apply
the same transformation to simulated levels. This affine transformation
preserves forecast rankings on the log-level scale; units are log percentage
points. The primary score is empirical marginal CRPS,
\begin{equation}
\widehat{\mathrm{CRPS}}(F,x)=\frac1M\sum_m|X^{(m)}-x|
-\frac1{2M^2}\sum_{m,l}|X^{(m)}-X^{(l)}|.
\end{equation}
The pairwise term is evaluated exactly from sorted draws. Supporting measures
are RMSE and MAE of sample means, and coverage and width of empirical 90\%
intervals. Width is interpreted alongside coverage and CRPS.

Stocks, origins, and selected endpoints receive equal weight. Each
engine/output/seed has 14,976 outcomes; the three-output comparison contains
539,136 score rows. Primary summaries average losses across the three fits.
RMSE takes the square root after averaging squared errors. This summarizes
performance across fits while retaining their common market outcomes.
CRPS skill is $100(1-\overline L_A/\overline L_B)$, with positive values
favoring A. Pooled skill is a ratio of mean losses. Stratum and stock-level
results show how those pooled effects vary across the panel.

Paired uncertainty uses 2,000 circular moving-block resamples of the 104
origins in blocks of 16. This spans the 160-observation forecast length at
ten-observation origin spacing. Indices are shared across stocks, horizons,
and fit seeds; fit losses are averaged within each origin. Percentile 95\%
intervals describe paired mean-CRPS differences conditional on the three
fitted models. The exploratory intervals are unadjusted for multiple
comparisons. Near/far skill differences describe horizon variation.

\clearpage
\section{Results}
\subsection{Incremental value across horizons}

\begin{table}[H]\centering\small
\caption{Mean marginal CRPS by horizon, averaging three fit losses. All columns use the same stocks and origins; lower is better.}
\begin{tabular}{lrrrrrr}\toprule Output & 5 & 10 & 20 & 40 & 80 & 160\\\midrule
EWMA, base & 3.802 & 5.792 & 9.700 & 17.474 & 33.480 & 64.814 \\
EWMA, Gaussian & 3.907 & 5.999 & 9.817 & 17.030 & 31.050 & 57.500 \\
EWMA, Flowz & 3.935 & 6.001 & 9.083 & 14.654 & 24.686 & 40.916 \\
ARIMA, base & 3.496 & 4.840 & 6.970 & 9.856 & 13.649 & 18.279 \\
ARIMA, Gaussian & 3.617 & 5.082 & 7.278 & 10.450 & 15.009 & 21.401 \\
ARIMA, Flowz & 3.970 & 5.722 & 8.190 & 11.388 & 16.244 & 22.902 \\
ARFIMA, base & 3.505 & 4.877 & 7.052 & 10.113 & 14.388 & 20.262 \\
ARFIMA, Gaussian & 3.682 & 5.272 & 7.770 & 11.625 & 17.530 & 25.940 \\
ARFIMA, Flowz & 4.066 & 6.024 & 8.866 & 13.013 & 19.724 & 29.126 \\
Structural, base & 3.480 & 4.826 & 6.955 & 9.881 & 13.801 & 18.516 \\
Structural, Gaussian & 3.606 & 5.067 & 7.254 & 10.457 & 15.121 & 21.423 \\
Structural, Flowz & 3.946 & 5.678 & 8.115 & 11.320 & 16.258 & 22.850 \\
\bottomrule\end{tabular}\label{tab:horizons}\end{table}

\begin{table}[H]\centering\small
\caption{CRPS skill by horizon band (\%). Positive values favor the first output. Difference is far minus near, in percentage points.}
\begin{tabular}{llrrr}\toprule Engine & Contrast & Near & Far & Difference\\\midrule
EWMA & Flowz/base & 1.4 & 30.7 & 29.2 \\
ARIMA & Flowz/base & -16.8 & -20.9 & -4.1 \\
ARFIMA & Flowz/base & -22.8 & -38.2 & -15.4 \\
Structural & Flowz/base & -16.2 & -19.5 & -3.3 \\
\bottomrule\end{tabular}\label{tab:bands}\end{table}

The result depends on the underlying baseline. Far-band Flowz-versus-baseline skills are EWMA: +30.7\%, ARIMA: -20.9\%, ARFIMA: -38.2\%, Structural: -19.5\%. Near-band skills are EWMA: +1.4\%, ARIMA: -16.8\%, ARFIMA: -22.8\%, Structural: -16.2\%. Positive skill is a reduction in mean CRPS. These comparisons concern the complete residual pipeline; the Gaussian control addresses how much of its behavior can be explained by location and scale alone.
Absolute scores place this improvement in context: far-band EWMA Flowz CRPS is 26.7520, while the ARIMA baseline achieves 13.9281. Thus EWMA provides the strongest incremental gain, while ARIMA supplies the lower absolute loss in this comparison.

\subsection{Value beyond Gaussian location and scale}

\begin{table}[H]\centering\small
\caption{Gaussian residual-control comparisons: CRPS skill (\%). Positive values favor the first output.}
\begin{tabular}{llrr}\toprule Engine & Band & G/base & Flowz/G\\\midrule
EWMA & Near & -2.2 & 3.6 \\
EWMA & Far & 8.8 & 24.0 \\
ARIMA & Near & -4.4 & -11.9 \\
ARIMA & Far & -12.1 & -7.8 \\
ARFIMA & Near & -8.4 & -13.4 \\
ARFIMA & Far & -23.1 & -12.3 \\
Structural & Near & -4.4 & -11.4 \\
Structural & Far & -11.4 & -7.3 \\
\bottomrule\end{tabular}\label{tab:control}\end{table}

\begin{table}[H]\centering\small
\caption{Incremental far-band Flowz-minus-Gaussian-control CRPS differences and exploratory 95\% origin-block intervals. Negative differences favor Flowz.}
\begin{tabular}{lrrr}\toprule Engine & Difference & Lower & Upper\\\midrule
EWMA & -8.441 & -9.212 & -7.726 \\
ARIMA & 1.225 & 0.930 & 1.506 \\
ARFIMA & 2.256 & 1.994 & 2.534 \\
Structural & 1.142 & 0.790 & 1.477 \\
\bottomrule\end{tabular}\label{tab:controluncertainty}\end{table}

Far-band Gaussian-control-versus-baseline skills are EWMA: +8.8\%, ARIMA: -12.1\%, ARFIMA: -23.1\%, Structural: -11.4\%. Flowz versus that control gives EWMA: +24.0\%, ARIMA: -7.8\%, ARFIMA: -12.3\%, Structural: -7.3\%. EWMA benefits from centering and scaling, followed by a larger improvement from conditional residual modeling. The paired far-band Flowz-minus-control interval lies below zero. For the parametric engines, the Gaussian correction worsens the baseline and Flowz increases that loss further.

\subsection{Variation across stocks and strata}
EWMA Flowz improves the Gaussian control in 24 of 24 stocks; stock-level skill ranges from 2.5\% to 35.3\%. ARIMA Flowz improves the Gaussian control in 6 of 24 stocks; stock-level skill ranges from -49.0\% to 4.7\%. ARFIMA Flowz improves the Gaussian control in 5 of 24 stocks; stock-level skill ranges from -64.1\% to 11.4\%. Structural Flowz improves the Gaussian control in 5 of 24 stocks; stock-level skill ranges from -48.0\% to 4.8\%. The stratum tables retain both horizon bands, showing how the pooled effects are distributed across the four selected cells.
Detailed stratum and stock-level results appear in Appendix~\ref{app:evidence}.

\subsection{Sensitivity to fit seed}

\begin{table}[H]\centering\small
\caption{Far-band CRPS skill by fit seed (\%). All fits use identical forecast seeds at each origin.}
\begin{tabular}{lrrr}\toprule Engine & Seed & Flowz/base & Flowz/G\\\midrule
EWMA & 123 & 29.8 & 23.0 \\
EWMA & 456 & 31.2 & 24.6 \\
EWMA & 789 & 31.0 & 24.4 \\
ARIMA & 123 & -21.3 & -8.2 \\
ARIMA & 456 & -20.3 & -7.3 \\
ARIMA & 789 & -21.2 & -8.0 \\
ARFIMA & 123 & -38.6 & -12.6 \\
ARFIMA & 456 & -36.9 & -11.2 \\
ARFIMA & 789 & -39.1 & -13.0 \\
Structural & 123 & -19.8 & -7.6 \\
Structural & 456 & -19.0 & -6.9 \\
Structural & 789 & -19.6 & -7.4 \\
\bottomrule\end{tabular}\label{tab:seeds}\end{table}

For Flowz versus baseline, far-band skill is positive for all three fits in EWMA; negative for all three fits in ARIMA, ARFIMA, Structural. For Flowz versus Gaussian control, far-band skill is positive for all three fits in EWMA; negative for all three fits in ARIMA, ARFIMA, Structural. The agreement in sign across all selected seeds makes the main engine-level pattern stable within this fitting experiment.

\section{Discussion}
The results identify a useful setting for conditional residual learning:
the drifting EWMA baseline leaves errors that Flowz can improve beyond
unconditional centering and scaling. With $\lambda_\mu=0.90$, the baseline
projects a short-memory drift estimate linearly over as many as 160
observations. One plausible explanation is that the conditional residual model
partly corrects this extrapolation. A conditional Gaussian or ridge residual
regression using the same state would help distinguish that mechanism from
the additional flexibility of the flow distribution.

The parametric engines yield a different result. Their baseline forecasts
have lower absolute far-band CRPS than the EWMA augmentation, and both the
Gaussian correction and conditional Flowz worsen them. This pattern makes
the choice of baseline central to the method's value. Examining which error
features persist across periods is a practical direction for selecting when
to apply a residual model.

The panel is selected from currently surviving stocks, and the relative
liquidity--volatility strata also differ in sector and other characteristics.
Stale closes can affect both apparent volatility and predictability. Overlapping
targets and shared market shocks reduce the effective information in the
104 origins. The study uses one configuration, three fit seeds, and a frozen
historical residual model. Rolling refits, independent periods and markets,
and conditional residual controls are the next empirical extensions.

\section{Conclusion}
Flowz learns a conditional distribution of direct multi-step baseline errors using random features and a fitted velocity field. Across three fits, it reduces far-band EWMA CRPS by 30.7\% against the baseline and 24.0\% against a Gaussian residual control with identical centering and scaling. The gain occurs at every selected seed and extends across all 24 stocks against that control. For ARIMA, ARFIMA, and structural baselines, the residual extension increases loss. The applied conclusion is that conditional residual learning can materially improve long-horizon forecasts when the baseline leaves predictable error structure; the implemented drifting EWMA provides the positive case in this panel.

Two directions follow from these results. The first is to identify other baselines whose residual structure can be improved by Flowz, using the EWMA result as a starting point. The second is to adapt the conditional flow to the ARIMA, ARFIMA, and structural baselines that did not improve in this experiment. Baseline-specific choices of conditioning features, model capacity, regularization, and training windows are natural candidates for this investigation, with tuning evaluated on chronological validation data and performance assessed on a subsequent held-out period.

\section*{Data and code availability}
The \texttt{flowz} R package, version 0.1.0, was published on CRAN on
October 18, 2026 \cite{flowzcran}:
\url{https://CRAN.R-project.org/package=flowz}.
Ancillary materials prepared for the arXiv submission include the scientific
source, frozen price panel, selection and corporate-action audits,
seed-averaged per-origin losses, per-seed summaries, and paired uncertainty
results. The compact companion provides environment versions, preserved source
hashes, and instructions for regenerating the full scores and fitted objects.
It omits bulk provider responses, fitted objects, and job logs. Market data
were obtained through Alpaca and remain subject to provider terms. The
evaluated implementation uses R 4.5.1, \texttt{forecast} 9.0.2, and
\texttt{fracdiff} 1.5-4, sourcing \texttt{flow\_engine.R},
\texttt{validation.R}, and \texttt{flow\_api.R}. The source is GPL-3.

\appendix
\section{Supporting experimental evidence}\label{app:evidence}
Base denotes the baseline, G the Gaussian residual control, and Flowz the
conditional residual model. A/B denotes CRPS skill of A relative to B; positive
values favor A. Paired differences use $L_A-L_B$, so negative values favor A.

\begin{table}[H]\centering\small
\caption{All-horizon scores, averaging losses across three fits. Errors, CRPS, and width are in log percentage points; coverage is a percentage for nominal 90\% intervals. Gaussian denotes the residual location--scale control.}
\begin{tabular}{lrrrrr}\toprule Output & RMSE & MAE & CRPS & Coverage & Width\\\midrule
EWMA, base & 56.362 & 28.849 & 22.511 & 68.9 & 57.965 \\
EWMA, Gaussian & 56.708 & 29.012 & 20.884 & 90.0 & 127.668 \\
EWMA, Flowz & 40.196 & 21.328 & 16.546 & 96.4 & 143.334 \\
ARIMA, base & 21.103 & 12.826 & 9.515 & 93.7 & 66.460 \\
ARIMA, Gaussian & 24.173 & 13.987 & 10.473 & 91.3 & 65.792 \\
ARIMA, Flowz & 26.462 & 15.235 & 11.403 & 94.5 & 80.948 \\
ARFIMA, base & 22.655 & 13.638 & 10.033 & 90.2 & 59.843 \\
ARFIMA, Gaussian & 26.103 & 15.034 & 11.970 & 91.7 & 87.957 \\
ARFIMA, Flowz & 29.427 & 17.115 & 13.470 & 94.6 & 107.446 \\
Structural, base & 22.742 & 12.923 & 9.577 & 92.9 & 64.907 \\
Structural, Gaussian & 25.189 & 14.002 & 10.488 & 90.3 & 64.232 \\
Structural, Flowz & 27.621 & 15.179 & 11.361 & 93.8 & 78.868 \\
\bottomrule\end{tabular}\label{tab:aggregate}\end{table}

\begin{table}[H]\centering\small
\caption{Far-band paired CRPS differences and exploratory 95\% origin-block intervals after averaging fit losses. Negative differences favor the first output.}
\begin{tabular}{llrrr}\toprule Engine & Contrast & Difference & Lower & Upper\\\midrule
EWMA & Flowz/base & -11.837 & -12.916 & -10.798 \\
EWMA & G/base & -3.396 & -3.865 & -2.970 \\
EWMA & Flowz/G & -8.441 & -9.212 & -7.726 \\
ARIMA & Flowz/base & 2.917 & 2.422 & 3.390 \\
ARIMA & G/base & 1.692 & 1.346 & 2.063 \\
ARIMA & Flowz/G & 1.225 & 0.930 & 1.506 \\
ARFIMA & Flowz/base & 5.699 & 4.935 & 6.474 \\
ARFIMA & G/base & 3.443 & 2.819 & 4.092 \\
ARFIMA & Flowz/G & 2.256 & 1.994 & 2.534 \\
Structural & Flowz/base & 2.743 & 2.300 & 3.190 \\
Structural & G/base & 1.601 & 1.248 & 2.019 \\
Structural & Flowz/G & 1.142 & 0.790 & 1.477 \\
\bottomrule\end{tabular}\label{tab:bootstrap}\end{table}

\begin{table}[H]\centering\scriptsize
\caption{Near-band CRPS skill by liquidity--volatility cell (\%). Positive values favor the first output.}
\begin{tabular}{llrrr}\toprule Stratum & Engine & G/base & Flowz/base & Flowz/G\\\midrule
high liquidity high volatility & EWMA & -0.5 & 4.8 & 5.3 \\
high liquidity high volatility & ARIMA & -3.2 & -10.4 & -7.0 \\
high liquidity high volatility & ARFIMA & -5.7 & -13.7 & -7.6 \\
high liquidity high volatility & Structural & -3.1 & -10.0 & -6.7 \\
high liquidity low volatility & EWMA & -0.9 & 2.5 & 3.3 \\
high liquidity low volatility & ARIMA & -1.5 & -9.3 & -7.7 \\
high liquidity low volatility & ARFIMA & -8.6 & -18.7 & -9.3 \\
high liquidity low volatility & Structural & -1.8 & -9.5 & -7.6 \\
low liquidity high volatility & EWMA & -5.9 & -6.0 & -0.1 \\
low liquidity high volatility & ARIMA & -8.0 & -28.7 & -19.1 \\
low liquidity high volatility & ARFIMA & -13.2 & -37.6 & -21.5 \\
low liquidity high volatility & Structural & -8.4 & -29.0 & -19.1 \\
low liquidity low volatility & EWMA & 0.2 & 8.4 & 8.2 \\
low liquidity low volatility & ARIMA & -2.0 & -13.0 & -10.8 \\
low liquidity low volatility & ARFIMA & -2.9 & -14.6 & -11.4 \\
low liquidity low volatility & Structural & -0.9 & -8.9 & -7.9 \\
\bottomrule\end{tabular}\label{tab:nearcohorts}\end{table}

\begin{table}[H]\centering\scriptsize
\caption{Far-band CRPS skill by liquidity--volatility cell (\%). Positive values favor the first output.}
\begin{tabular}{llrrr}\toprule Stratum & Engine & G/base & Flowz/base & Flowz/G\\\midrule
high liquidity high volatility & EWMA & 9.6 & 32.9 & 25.7 \\
high liquidity high volatility & ARIMA & -14.7 & -15.5 & -0.7 \\
high liquidity high volatility & ARFIMA & -19.6 & -22.6 & -2.5 \\
high liquidity high volatility & Structural & -12.0 & -13.2 & -1.1 \\
high liquidity low volatility & EWMA & 9.2 & 32.0 & 25.1 \\
high liquidity low volatility & ARIMA & -9.7 & -10.2 & -0.5 \\
high liquidity low volatility & ARFIMA & -22.6 & -30.4 & -6.3 \\
high liquidity low volatility & Structural & -10.3 & -10.9 & -0.5 \\
low liquidity high volatility & EWMA & 7.7 & 25.7 & 19.5 \\
low liquidity high volatility & ARIMA & -15.3 & -35.8 & -17.8 \\
low liquidity high volatility & ARFIMA & -35.1 & -69.5 & -25.5 \\
low liquidity high volatility & Structural & -16.3 & -36.8 & -17.7 \\
low liquidity low volatility & EWMA & 8.6 & 34.6 & 28.4 \\
low liquidity low volatility & ARIMA & -0.5 & -11.4 & -10.9 \\
low liquidity low volatility & ARFIMA & -3.1 & -11.7 & -8.4 \\
low liquidity low volatility & Structural & 1.6 & -3.9 & -5.6 \\
\bottomrule\end{tabular}\label{tab:farcohorts}\end{table}

\begin{table}[H]\centering\small
\caption{Far-band stock-level CRPS skill (\%), computed from each stock's seed-averaged losses. G denotes the Gaussian control.}
\begin{tabular}{llrrrr}\toprule Engine & Contrast & Median & Min. & Max. & Positive\\\midrule
EWMA & Flowz/base & 32.3 & 7.8 & 41.6 & 24/24 \\
EWMA & G/base & 8.8 & 2.8 & 13.1 & 24/24 \\
EWMA & Flowz/G & 25.8 & 2.5 & 35.3 & 24/24 \\
ARIMA & Flowz/base & -8.0 & -72.5 & 16.0 & 3/24 \\
ARIMA & G/base & -6.1 & -49.7 & 19.0 & 9/24 \\
ARIMA & Flowz/G & -4.8 & -49.0 & 4.7 & 6/24 \\
ARFIMA & Flowz/base & -24.3 & -249.4 & 23.3 & 5/24 \\
ARFIMA & G/base & -16.3 & -113.0 & 28.9 & 7/24 \\
ARFIMA & Flowz/G & -9.8 & -64.1 & 11.4 & 5/24 \\
Structural & Flowz/base & -7.4 & -80.3 & 4.3 & 4/24 \\
Structural & G/base & -1.4 & -51.4 & 9.7 & 10/24 \\
Structural & Flowz/G & -3.6 & -48.0 & 4.8 & 5/24 \\
\bottomrule\end{tabular}\label{tab:ticker}\end{table}

\subsection{Price-observation diagnostics}
All selected names have complete aligned sessions. Counts below cover the
1,196 evaluation sessions after December 31, 2021 through October 8, 2026.

\begin{table}[H]\centering\small
\caption{Evaluation-session counts of zero reported volume (Z) and unchanged split-adjusted closing price (U), by stock.}
\begin{tabular}{lrrlrr}\toprule Stock & Z & U & Stock & Z & U\\\midrule
AMAT & 0 & 0 & GROW & 0 & 116 \\
CAR & 0 & 1 & HAS & 0 & 6 \\
CBRL & 0 & 9 & HBCP & 0 & 11 \\
CHCI & 2 & 39 & JAKK & 0 & 5 \\
CHMG & 14 & 45 & JBHT & 0 & 2 \\
CLNE & 0 & 45 & NGNE & 0 & 18 \\
CZR & 0 & 13 & PNRG & 27 & 34 \\
CZWI & 5 & 59 & RARE & 0 & 6 \\
EXC & 0 & 9 & SSNC & 0 & 3 \\
FATE & 0 & 28 & TROW & 0 & 1 \\
GABC & 0 & 9 & VRDN & 0 & 13 \\
GBLI & 93 & 148 & YORW & 0 & 17 \\
\bottomrule\end{tabular}\label{tab:quality}\end{table}

\end{document}